\documentclass[runningheads]{llncs}

\usepackage[colorlinks,linkcolor=red,anchorcolor=blue, citecolor=blue]{hyperref}
\usepackage{multirow}
\usepackage[T1]{fontenc}
\usepackage{graphicx}
\usepackage{cite}
\usepackage{subfigure}
\usepackage{tabularx}
\usepackage{float}
\usepackage{amsmath} 
\usepackage{amssymb}
\usepackage{booktabs}
\usepackage{todonotes}
\usepackage{color}
\usepackage{overpic}
\usepackage{subfigure}
\usepackage{bm}
\usepackage{bbding}
\usepackage{makecell}
\usepackage{array}
\usepackage{caption}
\usepackage{bbding}
\usepackage{authblk} 

\usepackage[T1]{fontenc}
\usepackage{graphicx}
\begin{document}
\title{CIResDiff: A Clinically-Informed Residual Diffusion Model for Predicting Idiopathic Pulmonary Fibrosis Progression}
\titlerunning{CIResDiff: Prediction of Idiopathic Pulmonary Fibrosis Progression}

\author{Caiwen Jiang \inst{1,2} \and
Xiaodan Xing \inst{2} \and
Zaixin Ou \inst{1} \and
Mianxin Liu \inst{4} \and
Simon Walsh\inst{2} \and
Guang Yang \inst{2,5,7,8}\inst{(}\textsuperscript{\Envelope}\inst{)} \and
Dinggang Shen\inst{1,3,6}\inst{(}\textsuperscript{\Envelope}\inst{)}}

\authorrunning{Caiwen Jiang et al.}

\institute{
School of Biomedical Engineering \& State Key Laboratory of Advanced Medical Materials and Devices, ShanghaiTech University, Shanghai, China\\ 
\email{\{jiangcw,dgshen\}@shanghaitech.edu.cn}\\
\and
Bioengineering Department and Imperial-X, Imperial College London, London, UK\\
\email{g.yang@imperial.ac.uk}\\
\and 
Shanghai Clinical Research and Trial Center, Shanghai, 201210, China\\
\and
Shanghai Artificial Intelligence Laboratory, Shanghai 200232, China\\
\and
National Heart and Lung Institute, Imperial College London, London, UK\\
\and
Shanghai United Imaging Intelligence Co., Ltd., Shanghai, China\\
\and
Cardiovascular Research Centre, Royal Brompton Hospital, London, UK\\
\and
School of Biomedical Engineering \& Imaging Sciences, King's College London, London, UK\\
}
\begingroup
\renewcommand\thefootnote{}
\footnotetext{\Envelope\ Guang Yang and Dinggang Shen are co-senior last authors.}
\endgroup
\maketitle              
\begin{abstract}

The progression of Idiopathic Pulmonary Fibrosis (IPF) significantly correlates with higher patient mortality rates. Early detection of IPF progression is critical for initiating timely treatment, which can effectively slow down the advancement of the disease. However, the current clinical criteria define disease progression requiring two CT scans with a one-year interval, presenting a dilemma: \textit{a disease progression is identified only after the disease has already progressed}. To this end, in this paper, we develop a novel diffusion model to accurately predict the progression of IPF by generating patient's follow-up CT scan from the initial CT scan. Specifically, from the clinical prior knowledge, we tailor improvements to the traditional diffusion model and propose a Clinically-Informed Residual Diffusion model, called CIResDiff. The key innovations of CIResDiff include 1) performing the \textbf{target region pre-registration} to align the lung regions of two CT scans at different time points for reducing the generation difficulty, 2) adopting the \textbf{residual diffusion} instead of traditional diffusion to enable the model focus more on differences (i.e., lesions) between the two CT scans rather than the largely identical anatomical content, and 3) designing the \textbf{clinically-informed process} based on CLIP technology to integrate lung function information which is highly relevant to diagnosis into the reverse process for assisting generation. Extensive experiments on clinical data demonstrate that our approach can outperform state-of-the-art methods and effectively predict the progression of IPF.

\end{abstract}
\section{Introduction}
Idiopathic Pulmonary Fibrosis (IPF) is a severe and irreversible lung disease that scars and thickens lung tissues, leading to respiratory difficulties~\cite{spagnolo2021idiopathic, maher2021global}. Timely treatment of IPF can effectively slow down its process and improve patients' quality of life~\cite{torrisi2020evolution, finnerty2021efficacy}. The progression of IPF can either remain stable or exacerbate over time. Consequently, to save healthcare expenses, in some countries (especially those with universal healthcare like the UK), antifibrotic treatment is initiated only if the IPF is confirmed to exacerbate over time. This approach presents a challenge: treatment is delayed until the fibrosis has advanced, losing early intervention opportunities for patients at high risk but not yet showing significant progression. In this context, predicting the progression of IPF in advance is extremely important for enabling timely treatment and reducing healthcare costs.

A feasible approach for predicting the progression of IPF is to generate the follow-up CT scan from the initial CT scan. This method of disease prediction or diagnosis through generation has already achieved success in numerous studies~\cite{han2022image, jiang2023s2dgan, frid2018gan, ma2020combining, jiang2024real}. For instance, Han \textsl{et al} adopt the regularized generative adversarial networks to generate images of future time points for predicting the risk of osteoarthritis~\cite{han2022image}. Jiang \textsl{et al} employ a transformer-based generative adversarial network to generate dual-energy CT images from single-energy CT for diagnosing postoperative cerebral hemorrhage~\cite{jiang2023s2dgan}.  Moreover, 
we choose to generate the follow-up CT scan rather than directly predicting disease progression from the initial CT scan considering the following facts: 1) The use of follow-up CT scan is more clinically natural in terms of diagnosis and thus more subjectively convincing. 2) The generation of follow-up CT scan can better exploit in the information entailed initial CT scan, thus more likely to produce a correct diagnosis. 3) The generated follow-up CT scan can provide additional information, such as the location of the lesion area.

Among the existing image generation techniques, the diffusion model has shown large potential and obtained great success~\cite{ho2020denoising,sohl2015deep,yue2024resshift}. They accomplish this by converting complex generation tasks into a series of simpler denoising tasks, enabling more stable and detailed generation. Meanwhile, clinical observations reveal the following prior information. \textbf{First}, due to patient posture and physiological movement, there is a significant spatial difference between the two CT scans. \textbf{Second}, as both scans are from the same individual, the majority of their content (i.e., the anatomical structure) is identical, while the lesion areas we focus on only constitute a minor part. \textbf{Lastly}, patients typically undergo lung function tests during their initial CT scan, and the information from these tests is highly relevant to the progression of IPF.

Taking all into consideration, in this paper we propose a Clinically-Informed Residual Diffusion model (CIResDiff) based on the traditional diffusion model to predict the progression of IPF. Specifically, we improve the traditional diffusion model by 1) performing target region pre-registration for the two CT scans to enable voxel-to-voxel generation for reducing the complexity, 2) adopting residual diffusion to make the model's generation process focus more on the differences (i.e., the lesion areas) between the two CT scans for more precise generation of target areas and fast model inference, and 3) designing a clinically-informed process based on CLIP~\cite{radford2021learning} technology to capture the correlation between lung function test information and corresponding CT scans, and utilizing this correlation to guide the generation for producing images with higher diagnostic value.

The main contributions of our work include: \romannumeral1) the first attempt to predict the progression of IPF by generating follow-up CT scans from initial scans, \romannumeral2) developing a novel diffusion model (CIResDiff) tailored for this task based on clinical prior knowledge, and \romannumeral3) demonstrating the effectiveness of our CIResDiff on collected datasets by extensive experiments.

\begin{figure}[!t]
\setlength{\belowcaptionskip}{-0.4cm}
\centering
\begin{overpic}[width=1\linewidth]{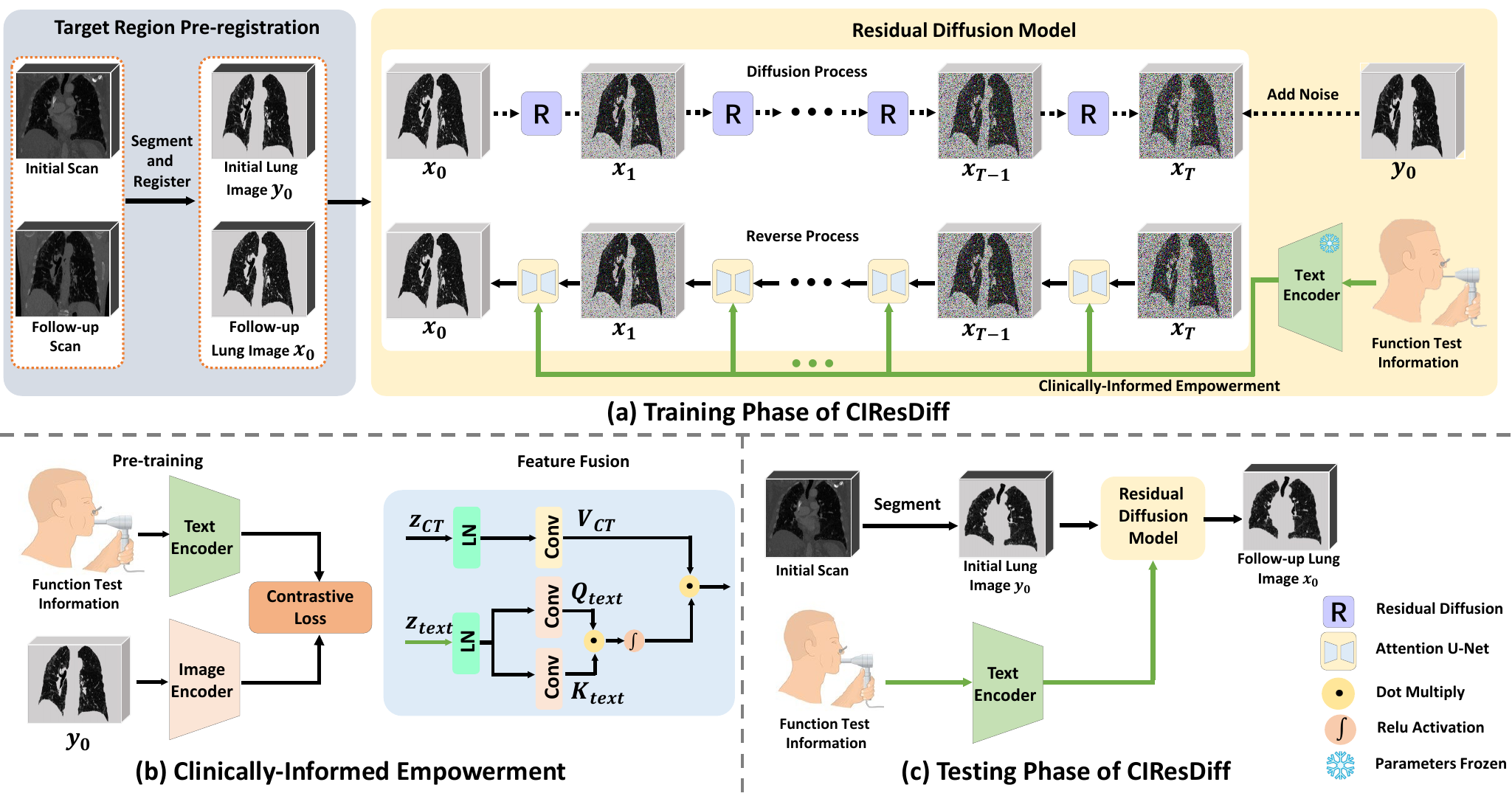}

    \end{overpic}
    \vspace{-1mm}
\centering
 \setlength{\abovecaptionskip}{0.1cm}
 
\caption{Overview of proposed CIResDiff. (a) and (c) provide the framework of CIResDiff as well as depict its implementation during both the training and testing phases, and (b) illustrates the details of clinically-informed process.}
\label{framework}
\end{figure}

\section{Method}
Our method is illustrated in Fig.~\ref{framework}. First, we extract and align the lung region from the two CT scans by the target region pre-registration. Then, in the residual diffusion process, the follow-up lung image $x_{0}$ is converted to the noisy initial lung image $x_{T}$ by incrementally incorporating the differences of two CT scans. Subsequently, the noisy initial lung image $x_{T}$ undergoes a series of reconstruction steps guided by the clinically-informed process until it's converted back to the follow-up lung image $x_{0}$. Note that only the reverse process is involved during the testing phase. In the following, we introduce the details of target region pre-registration, residual diffusion, and CLIP-based text processing module.

\subsection{Target Region Pre-registration}

Voxel-to-voxel supervised generation is less challenging compared to direct unsupervised generation. Hence, we employ a target area pre-registration to obtain spatially-aligned lung image pairs from the initial and follow-up CT scans for model training.

Specifically, for the input of two CT scans, to eliminate background interference, we first use the TotalSegmentator~\cite{wasserthal2023totalsegmentator}, an open-access tool based on nnU-Net and trained with more than one thousand samples, to segment the left and right lung regions from both CT scans. Then, we perform a dilation operation on the segmented results to preserve lung-surrounding tissues relevant to diagnosis. Subsequently, we apply an affine registration method~\cite{avants2009advanced} to align the segmented left and right lung regions individually. Finally, we merge the aligned left and right lung regions to obtain spatially-aligned lung region image pairs for model training.

\subsection{Residual Diffusion}
In the process of generating follow-up scans from initial scans, the main learning objective for the model is the lesion areas that are crucial for diagnosis but occupy a small proportion. Thus, we integrate residual learning with the diffusion model to enhance the model's ability to learn changes in such lesion areas, proposing a residual diffusion strategy. This strategy enables the diffusion model to exclusively learn the residuals between the initial and follow-up scans.

Unlike the traditional diffusion process, which disrupts the input initial lung image $x_0$ into pure Gaussian noise, our proposed residual diffusion converts the input initial lung image $x_0$ into noisy follow-up lung image $x_T$. Specifically, we first calculate the difference $e_0 = y_0 - x_0$ between the initial lung image $y_0$ and the corresponding follow-up lung image $x_0$, and then apply a shifting sequence $\{\eta_t\}_{t=1}^T$ to incrementally add this difference $e_0$ to $x_0$. In this way, the residual diffusion process can be formulated as follows: 
\begin{equation}
q(x_t | x_0, y_0) = \mathcal{N}(x_t; x_0 + \eta_t e_0, k^2 {\eta}_t I),     \quad t=1,2,...,T,
\label{equation 1}
\end{equation}
where $t$ is the timestep which satisfies ${\eta_1}$ $\rightarrow$ 0, ${\eta_T}$ $\rightarrow$ 1, $k$ is a hyper-parameter controlling the noise variance, and $I$ is the identity matrix.

In this context, the reverse process, corresponding to the residual diffusion process, involves recovering the follow-up lung image from the noisy initial lung image rather than from pure noise. Such reverse process can estimate the posterior distribution $p(x_0|y_0)$ via the following formulation: 

\begin{equation}
\begin{split}
 p(x_{0}|y_{0}) &= \int p(x_{T}|y_{0})\prod_{t=1}^{T}p_{\theta}(x_{t-1}|x_{t},y_{0})\mathrm{d}x_{1;T},\\
 p_{\theta}(x_{t-1}|x_{t},&y_{0}) = \mathcal{N}(x_{t-1};\mu_{\theta}(x_{t},y_{0},t),\Sigma_{\theta}(x_{t},y_{0},t)),\\
  &p(x_{T}|y_{0}) \approx \mathcal{N}(x_T; y_0, k^2I),
 \end{split}
\label{equation 2}
\end{equation}
where $\Sigma_{\theta}(x_{t},y_{0},t)$ is a fixed variance and $\mu_{\theta}(x_{t},y_{0},t)$ can be reparameterized as follows:
\begin{equation}
\mu_{\theta(x_{t},y_{0},t)}=\frac{\eta_{t-1}}{\eta_{t}}x_{t}+\frac{\eta_t-\eta_{t-1}}{\eta_{t}}f_{\theta}(x_{t},y_{0},t).
\label{equation 3}
\end{equation}

Here, $f_{\theta}$ is a deep neural network with parameter $\theta$, aiming to $\hat{x}_0$, In our implementation, we utilize the attention U-Net~\cite{oktay2018attention} as $f_{\theta}$. The reverse process can be trained through the following reconstruction loss:
\begin{equation}
\mathcal{L}_{rec} = \min_{\theta}\sum_{t}\left\|f_{\theta}(x_{t},y_{0},t)-x_0\right\|_2^2
\label{equation 4}
\end{equation}

\subsection{Clinically-informed Process }
During the diagnosis of IPF, lung function tests are typically conducted in addition to two CT scans. The information derived from these tests is highly relevant to the lung anatomical structure. Therefore, we believe that capturing the correlation between function test information and the corresponding CT scan, and using it to guide the generation process, can aid in producing CT scans with higher diagnostic value. To achieve this, we design a clinically-informed process based on Contrastive Language-Image Pretraining (CLIP)~\cite{radford2021learning} technology. 

Details of the clinically-informed process are shown in Fig.~\ref{framework} (b), which includes pre-training and feature fusion stages. In the pre-training stage, we extract textual and image features from function text information and the corresponding initial scan using text and image encoders, respectively. Then, we constrain these features to be aligned using the contrastive loss. In this way, we can obtain the pre-trained text encoder for incorporating function text information into the reverse process. Note that the samples used for pre-training are exclusively from the training set and do not involve any test samples.

In the feature fusion stage, for a particular reconstruction in the reverse process, the pre-trained text encoder first extracts textual features $z_{text}$ from function test information. Then $z_{text}$ is fed into a denoising attention U-Net~\cite{oktay2018attention} at each step for calculation of cross-attention, where the query $Q$ and key $K$ are calculated from $z_{text}$ while the value $V$ is still calculated from the output of the previous layer because our final goal is CT estimation. Denoting the output of previous layer as $z_{CT}$, the CT-guided cross-attention can be formulated as follows:

\begin{equation}
\begin{split}
& Output = softmax(\frac{Q_{text}K_{text}^T}{\sqrt{d}}+B) \cdot V_{PET},\\
 Q_{text} = Con&v_Q(z_{text}),\quad K_{text} = Conv_K(z_{text}),\quad V_{CT} = Conv_V(z_{CT}),
 \end{split}
\label{equation 2}
\end{equation}
where $d$ is the number of channels, $B$ is the position bias, and $Conv(\cdot)$ denotes the $1 \times 1 \times 1$ convolution with stride of 1.

\section{Experiments}

\subsection{Dataset and Implementation}
Our dataset originates from OSIC\footnote{\url{https://www.osicild.org}}, an openly accessible global database containing numerous CT scans related to Idiopathic Pulmonary Fibrosis (IPF). We collect a total of 200 samples, each comprising of two CT scans with a $46.3\pm7.8$ weeks interval, along with corresponding lung function test information (including physiological indicators such as vital capacity, peak expiratory flow, etc.) and diagnostic labels. Among these 200 samples, 160 samples are used for training and the remaining 40 samples are used for testing. During the evaluation, we conduct five-fold cross-validation to exclude randomness.

In our implementation, experiments were conducted on the PyTorch platform using two NVIDIA Tesla A100 GPUs and an Adam optimizer with initial learning rate of 0.001. All images are resampled to voxel spacing of $1\times1\times1~\text {mm}^{3}$ and resolution of $512 \times 512 \times 256$, while their intensity range is normalized to [$0,1$] by min-max normalization. For increasing the training samples and reducing the dependence on GPU memory, we extract the overlapped patches of size $96 \times 96 \times 96$ from every whole CT scan. We evaluate the quantitative results by two commonly used quantitative metrics, including Peak Signal to Noise Ratio (PSNR)~\cite{jiang2023semi} and Structural Similarity Index (SSIM).

\begin{figure}[!t]
	\begin{minipage}{0.5\linewidth}
    \centering
\resizebox{60mm}{14mm}{\begin{tabular}{l|c|c}
\toprule[0.5mm]
 Method    &  \multicolumn{1}{c|}{PSNR [dB]$\uparrow$ } & \multicolumn{1}{c}{SSIM [\%]$\uparrow$}   \\ 
\hline
DM      &         $25.84~\pm~0.93$   &  $98.19~\pm~ 1.24$     \\
DM-CIP     &       $26.12 ~\pm~ 1.07$ & $98.75~\pm~ 1.44$  \\
DM-R    &       $26.84 ~\pm~ 0.92$ & $99.03~\pm~ 1.16$  \\
DM-R-CIP  &\bm{$27.52~\pm~0.83$} & \bm{$99.25~\pm~ 0.92$} \\   [3pt]   \bottomrule[0.4mm]              
\end{tabular}}
        \captionof{table}{Quantitative results of ablation analysis, in terms of PSNR and SSIM.}
        \label{table1}
\end{minipage}
	\hfill
	\begin{minipage}{0.5\linewidth}
		\centering
   
		\centering
		\vspace{-0.6cm}
		\setlength{\abovecaptionskip}{0.28cm}
		\includegraphics[width=\linewidth]{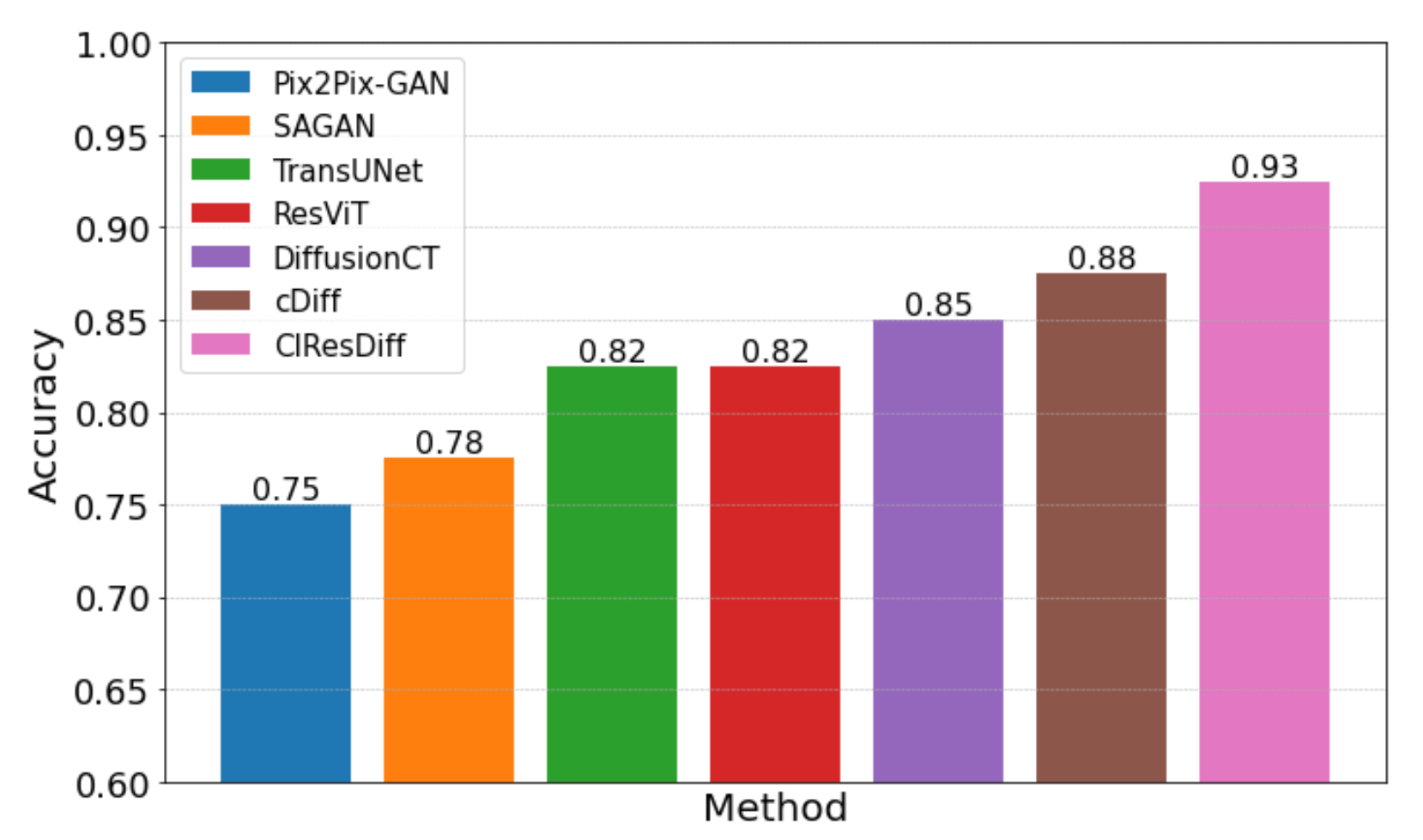}
		\vspace{-0.6cm}
		\caption{Diagnostic evaluation.}
		\label{Diagnostic}

	    \end{minipage}
\end{figure}

\subsection{Ablation Analysis}
To verify the effectiveness of our proposed strategies, i.e., \textsl{residual diffusion} and \textsl{clinically-informed process}, we design another four variant diffusion models (DMs) including: 1) DM: standard DM; 2) DM-CIP: DM with clinically-informed process; 3) DM-R: DM with residual diffusion; 4) DM-R-CIP: DM with residual diffusion and clinically-informed process. All methods use the same experimental settings, and their quantitative results are given in Table~\ref{table1}.

The quantitative results are provided in Table~\ref{table1}, from which, we can find the following observations.
(1) DM-CIP with the clinically-informed process achieves better performance than DM. This proves that integrating highly relevant clinical information (i.e., function test information) into the reverse process is beneficial for the generation of follow-up CT scans. (2) DM-R, employing residual diffusion, achieves better performance than DM employing the traditional diffusion process, proving that residual diffusion is more appropriate for learning the mapping between highly similar initial and follow-up CT scans. (3) DM-R-CIP achieves better results than all other variants on both PSNR and SSIM, which shows both of our proposed strategies contribute to the final performance. These three comparisons conjointly verify the effective design of our proposed CIResDiff, where the \textsl{residual diffusion} and \textsl{clinically-informed process} both benefit our generation task.

\begin{table}[!t]
 \setlength{\abovecaptionskip}{0.1cm}
\setlength{\belowcaptionskip}{0.1cm}
\centering
\renewcommand\arraystretch{1.3}
\setlength\tabcolsep{8pt}
\caption{Quantitative comparison of our CIResDiff with several state-of-the-art generation methods, in terms of PSNR and SSIM.}
\resizebox{90mm}{20mm}{
\begin{tabular}{l|c|c} 
\toprule[0.5mm]
Method         & \multicolumn{1}{c|}{PSNR [dB]$\uparrow$ } & \multicolumn{1}{c}{SSIM [\%]$\uparrow$} \\ 
\hline
Pix2Pix-GAN~\cite{isola2017image}                &$22.53~\pm~2.13$  & $94.16~\pm~1.59$   \\
SAGAN~\cite{lan2021three}                       &$22.93~\pm~1.98$  & $94.84~\pm~ 1.44$   \\
\hline
TransUNet~\cite{chen2021transunet}               &$23.35~\pm~1.72$  & $95.49~\pm~ 1.24$   \\
ResViT~\cite{dalmaz2022resvit}                   &$23.62~\pm~1.65$  & $95.96~\pm~1.32$   \\
\hline
DiffusionCT~\cite{selim2023diffusionct}        &$25.21~\pm~0.92$    & $97.82~\pm~ 1.21$    \\
cDiff~\cite{peng2023cbct}                       &$25.84~\pm~0.93$   & $98.19~\pm~ 1.24$    \\
CIResDiff                                       &\bm{$27.52~\pm~0.83$} & \bm{$99.25~\pm~ 0.92$}   \\   
\bottomrule[0.4mm]
\end{tabular}}
\label{table2}
\end{table}

 \begin{figure}[!t]
 \setlength{\abovecaptionskip}{0.1cm}
\setlength{\belowcaptionskip}{-0.4cm}
\centering
\begin{overpic}[width=1\linewidth]{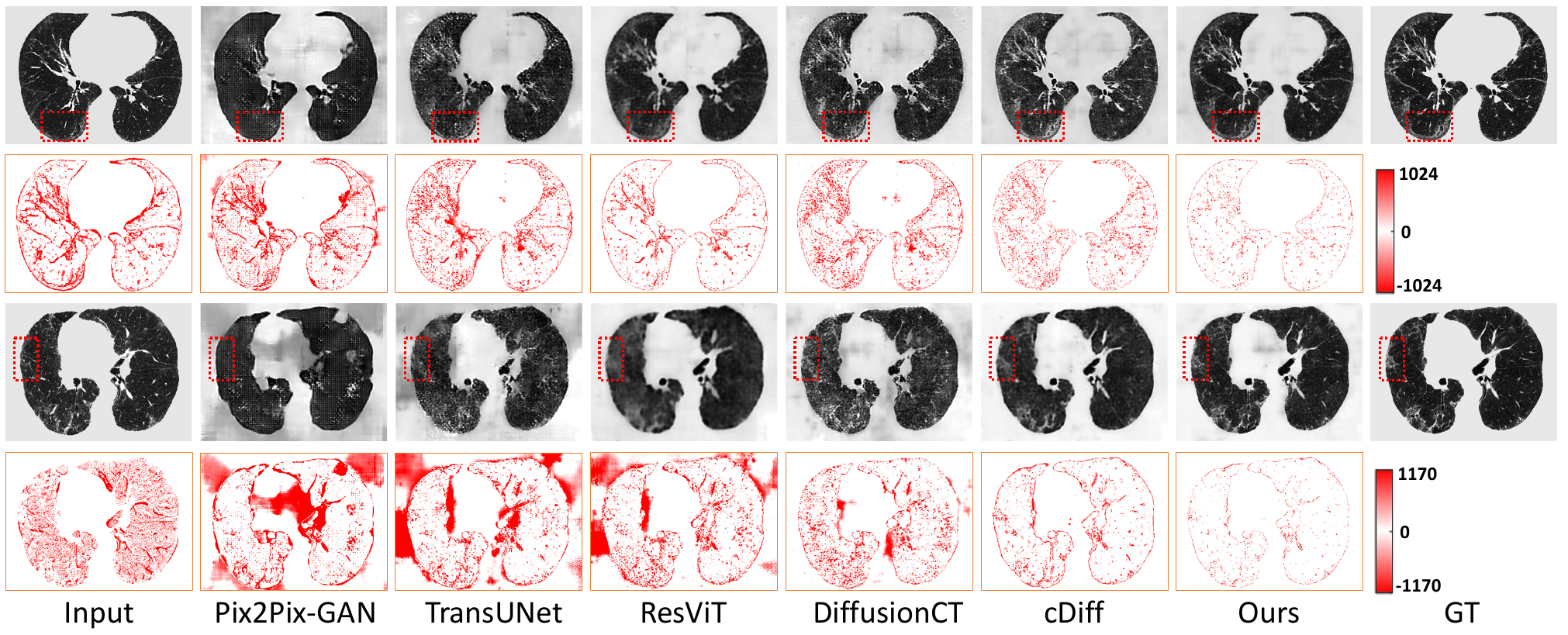}
        
    \end{overpic}
    \vspace{-4mm}
\centering
\caption{Visual comparison of follow-up lung images produced by six different methods. From left to right are the input (initial scan), results of five other comparison methods (2nd-6th columns) and our CIResDiff (7th column), and the ground truth (follow-up scan). The corresponding difference maps between the generated results and GT are shown in the 2nd and 4th rows, where darker colors indicate larger differences. Red boxes show the lesion areas for detailed comparison.}
\label{fig3}
\end{figure}

\subsection{Comparison with State-of-the-art Methods}
We further compare our CIResDiff with six state-of-the-art generation methods, which can be divided into three classes: 1) GAN-based methods, including Pix2Pix-GAN~\cite{isola2017image} and SAGAN~\cite{lan2021three}; 2) transformer-based methods, including TransUNet~\cite{chen2021transunet} and ResViT~\cite{dalmaz2022resvit}; 3) diffusion model-based methods, including DiffusionCT~\cite{selim2023diffusionct} and conditional Diffusion (cDiff)~\cite{peng2023cbct}. The quantitative and qualitative results are provided in Table~\ref{table2} and Fig.~\ref{fig3}, respectively.

\textsl{Quantitative Comparison:} 
The quantitative results are provided in Table~\ref{table2}. From the table, it is evident that diffusion model-based methods generally outperform GAN-based and transformer-based methods, validating our selection of the diffusion model as the baseline. Moreover, among all diffusion model-based methods, our CIResDiff achieves the optimal results, with improvements in PSNR and SSIM over the sub-optimal cDiff by $1.68$ dB and $1.06 \%$, respectively. This demonstrates the effectiveness of our targeted improvements to the traditional diffusion model, including residual diffusion and clinically-informed process.

\textsl{Qualitative Comparison:}
We provide a visual comparison of follow-up lung images generated by six different methods in Fig.~\ref{fig3}. First, compared to other methods, our CIResDiff can generate the overall optimal images, characterized by the least noise, fewest artifacts but clearest structure. Second, in terms of detail, our CIResDiff can also most accurately generate the lesion areas (i.e., areas marked by red boxes) that are crucial for predicting the progression of IPF. Finally, the lightest color in the difference map demonstrates our CIResDiff can generate lung images with the smallest difference from the ground truth. Such key observations demonstrate that our CIResDiff is superior to those state-of-the-art methods.

\vspace{-4mm}
\subsection{Diagnostic Evaluation}
Our ultimate goal is to predict the progression of IPF using generated follow-up lung images. Therefore, we design relevant downstream diagnostic tasks to assess the diagnostic value of follow-up lung images generated by different methods. Specifically, we first train a ResNet-based classifier using real image pairs (i.e.,  real follow-up lung image and real initial lung image) from the training set. During the training process, follow-up and initial lung images are concatenated together as input to predict diagnostic labels. Then, we use the pre-trained classifier to assess the diagnostic value of images generated by different methods. During evaluation, the input is the fake image pairs (i.e., generated follow-up lung image and real initial lung image). The prediction results are provided in Fig.~\ref{Diagnostic}.

Fig.~\ref{Diagnostic} shows that the follow-up lung images generated by our CIResDiff yield the best results for predicting IPF compared to other methods. Specifically, compared to Pix2Pix-GAN, which obtains the worst results, our CIResDiff shows a significant improvement in accuracy by $18\%$. This may be due to our method's incorporation of highly relevant lung function information into the generation process, thereby producing images more beneficial for diagnosis. These results further confirm the effectiveness of using generation for IPF prediction and highlight the clinical application potential of our CIResDiff.

\section{Conclusion}
In this paper, to achieve early prediction of IPF progression, we develop a novel diffusion model named CIResDiff to predict patients's follow-up CT scan from the initial CT scan. To facilitate more precise and effective generation, our CIResDiff employs three strategies: 1) pre-aligning lung regions in both CT scans to reduce the complexity of generation; 2) learning residuals between initial and follow-up CT scans to focus more on lesion areas critical for prediction; 3) integrating clinical lung function test information to help generate results with greater diagnostic value. Extensive experiments demonstrate that our method outperforms state-of-the-art approaches and can generate images with higher diagnostic value.

\subsubsection{\ackname} This work was supported in part by National Natural Science Foundation of China (grant numbers U23A20295, 62131015, 62250710165), the STI 2030-Major Projects (No. 2022ZD0209000), Shanghai Municipal Central Guided Local Science and Technology Development Fund (grant number YDZX20233100001001), the China Ministry of Science and Technology (STI2030-Major Projects-2022ZD0213100), The Key R\&D Program of Guangdong Province, China (grant numbers 2023B0303040001, 2021B0101420006), the ERC IMI (10100\\5122), the H2020 (952172), the MRC (MC/PC/21013), the Royal Society (IEC\textbackslash NS\\FC\textbackslash211235), the NVIDIA Academic Hardware Grant Program, the SABER project supported by Boehringer Ingelheim Ltd, NIHR Imperial Biomedical Research Centre (RDA01), Wellcome Leap Dynamic Resilience, UKRI guarantee funding for Horizon Europe MSCA Postdoctoral Fellowships (EP/Z002206/1), and the UKRI Future Leaders Fellowship (MR/V023799/1). This work was completed under the close collaboration between Caiwen Jiang and Xiaodan Xing, and they contributed equally to this work.

\subsubsection{Declaration of Competing Interest.}
The authors declare that they have no known competing financial interests or personal relationships that could have appeared to influence the work reported in this paper.

{\small 
\bibliographystyle{splncs04} 
\bibliography{myref.bib} }

\end{document}